\documentclass[a4paper,11pt]{article}
\usepackage{pos}
\usepackage{tabularx}
\usepackage{booktabs}
\title{Simulations of high-energy neutrino emissions from blazars with the LeHa-Paris code}
\author*[a,b]{Francesco Carenini}
\author[c]{Matteo Cerruti}

\affiliation[a]{Università di Bologna, Dipartimento di Fisica e Astronomia\\
Viale Berti Pichat 6/2, 40127 Bologna, Italia}

\affiliation[b]{INFN - Sezione di Bologna\\
Viale Berti Pichat 6/2, 40127 Bologna, Italia}

\affiliation[c]{Université Paris Cité, Astroparticule et Cosmologie (APC)\\
10 Rue Alice Domon et Léonie Duquet, 75013 Paris, France}

\emailAdd{francesco.carenini2@unibo.it}
\emailAdd{cerruti@apc.in2p3.fr}

\abstract{
The identification of astrophysical sources responsible for high-energy cosmic neutrinos has long been a challenge. A significant milestone was achieved with the blazar TXS 0506+056, which was found to be in a flaring state of high gamma-ray emission and associated at the 3sigma level with a 290 TeV neutrino detected by IceCube in September 2017. This discovery motivated deeper exploration of the theoretical link between photon and neutrino emissions. In this context, simulations of proton-photon interactions in blazars and radiative processes are conducted using advanced numerical codes to predict neutrino spectra. The LeHa-Paris code, previously applied to TXS 0506+056, enables the computation of both leptonic and hadronic components of blazar spectral energy distributions, facilitating exploration of a broad parameter space. In this work, starting from the case of PKS 2155-304, one of the brightest and most studied High-frequency-peaked BL Lacs (HBLs), known for its extreme variability and subject of multi-wavelength observational campaigns, a methodology has been developed to extend neutrino flux templates, optimized via LeHa-Paris, to the full class of HBLs. Afterwards, neutrino emission models for a subset of HBLs from the 3HSP catalogue are derived.}

\ConferenceLogo{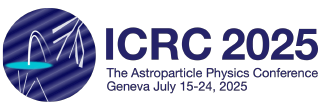}

\FullConference{39th International Cosmic Ray Conference (ICRC2025)\\
 15–24 July 2025\\
Geneva, Switzerland\\}
\begin{document}
\maketitle

\section{Introduction}

Relativistic jets launched by accreting supermassive black holes, known as active galactic nuclei (AGN), radiate across the entire electromagnetic spectrum, from radio waves to TeV gamma rays. Due to Doppler boosting, AGNs with jets aligned toward the observer, classified as blazars, appear exceptionally bright. Blazars exhibit non-thermal spectral energy distributions (SEDs) with two distinct components. The low-energy component peaks between the infrared and X-ray bands and is well understood as synchrotron radiation from relativistic electrons and positrons in the jet. The high-energy component, peaking in the gamma-ray range from MeV to TeV, is less well understood.\\
In leptonic models, this high-energy emission is attributed to inverse Compton (IC) scattering by the same electron/positron population responsible for the synchrotron emission. This scattering may occur on synchrotron photons themselves, known as the synchrotron self-Compton (SSC) process, or on external photon fields near the black hole. While hadrons (protons and nuclei) may be present in the jet, they do not significantly contribute to the emission in leptonic scenarios.\\
Hadronic models, by contrast, explain the high-energy component via processes involving relativistic protons, either through direct synchrotron emission or through interactions with photons that produce secondary leptons. While both leptonic and hadronic models often yield similar photon spectra, a key distinction lies in neutrino production: leptonic models do not predict neutrinos, whereas hadronic models do.\\
The identification of blazar TXS 0506+056 as a potential cosmic neutrino source \cite{IceCube:2018dnn} has elevated the significance of blazars as key candidates for neutrino emission. This discovery has reignited interest in hadronic blazar models. Supporting this, the LeHa-Paris code \cite{Cerruti:2014iwa}, already applied to TXS 0506+056 \cite{Cerruti:2019mnras}, enables the computation of both leptonic and hadronic components of blazar SEDs. Starting with the well-known and extensively studied PKS 2155-304, a prominent High-frequency-peaked BL Lac (HBL), a methodology has been developed to use LeHa-optimized models to generate neutrino flux templates that can be extended to the entire HBL class.  This study is particularly timely, as lepto-hadronic models can be directly tested against neutrino observations from Cherenkov neutrino telescopes, such as the ones under construction in the Mediterranean Sea by the KM3NeT Collaboration \cite{KM3Net:2016zxf}. 

\section{LeHa-Paris code}

LeHa-Paris is a steady-state numerical model developed to compute photon and neutrino emission from populations of relativistic electrons and protons in equilibrium within a spherical emitting region inside a relativistic jet. Specifically, this work has adopted a one-zone emission model, which assumes a spherical emitting region within the jet characterized by a radius \(R\), magnetic field strength \(B\), and Doppler factor \(\delta\). This is the sole region assumed to account for the observed emission. Although radio emission is typically self-absorbed in this region and thus underestimated, one-zone models effectively reproduce the broad-band SEDs and variability patterns observed in blazars by attributing the dominant emission at any given time to this single zone. In their simplicity, one-zone models have been successful in explaining the emission from blazars, both for the broad-band SEDs, and for the time variability and multi-wavelength correlations.\\
\\
The primary electron population is described by a broken power-law distribution parameterized by the minimum, break and maximum Lorentz factors \(\gamma_{\min}\), \(\gamma_{\mathrm{break}}\), \(\gamma_{\max}\), the spectral indices \(\alpha_{e,1}\) and \(\alpha_{e,2}\), and a normalization constant \(K_e\). Internal absorption due to \(\gamma\gamma\) pair production is also implemented. For secondary leptons coming from \(\gamma\gamma\) interactions, their steady state distributions are computed from injection, synchrotron and IC cooling and adiabatic losses.\\
The hadronic component of the model calculates proton synchrotron radiation similarly to the electron case and simulates proton-photon (\(p\gamma\)) interactions, assuming that the target photon field comprises synchrotron photons from both primary electrons and protons, as well as SSC photons. The energy distributions of secondary particles injected via photo-meson interactions are derived by running the Monte Carlo code SOPHIA \cite{Mucke2000}, which has been modified to accept arbitrary target photon fields, extract muon spectra, and include synchrotron cooling of kaons, pions, and muons. These SOPHIA outputs serve as injection terms to compute the steady-state distributions of secondary particles. Photon and neutrino spectra are obtained by multiplying their injection rates by their respective escape timescales.\\
Pair injection from Bethe–Heitler pair production is computed, with the steady-state distribution calculated analogously to other leptons. Bethe–Heitler processes have been shown to play a significant role in BL Lac objects by producing a secondary population of electron-positron pairs that contribute emission in the intermediate energy range between X-rays and gamma rays \cite{Petropoulou2015}, effectively filling the gap in the SED. Pair cascades are computed iteratively: the code first calculates the steady-state leptonic distribution along with its synchrotron and IC radiation, then computes pair injection from \(\gamma\gamma\) pair production, determines its steady state and associated radiative emission, and continues this iterative process up to the fifth cascade generation. The contribution from higher generations is considered negligible. The underlying assumption is that the cascade is never self-sustained but always driven by synchrotron and SSC photons from primary electrons or by synchrotron photons from primary protons.
\\
In recent efforts \cite{2024arXiv241114218C}, the code was compared against multiple independent hadronic blazar codes, including AM$^3$, B13, ATHE$\nu$A, and LeHaMoC. These comparisons demonstrated close agreement in the shapes of photon and neutrino spectra, proving that the results are independent from the specific radiative code used in the simulation. Absolute flux normalizations differed by up to $\sim$40\%, consistent with implementation-dependent variations in interaction cross sections and particle tracking schemes.

\section{The case of PKS 2155-304}
PKS 2155-304, located at redshift \( z = 0.117 \), with right ascension \( 21^\text{h} 58^\text{m} 52.0^\text{s} \) and declination \(-30^\circ 13' 32''\), is a well known southern object classified as an HBL. PKS 2155-304 has been the focus of extensive study, with several authors noting potential contamination of the hard X-ray spectra by a high-energy component, referred to as the hard tail \cite{Madejski:2016evb}. However, the absence of very high-energy (VHE, \( E > 100\,\text{GeV} \)) data during these studies limited constraints on the VHE \(\gamma\)-ray flux. A previous multi-wavelength campaign, using X-ray instruments, Fermi-LAT, and H.E.S.S., successfully modeled the data with both leptonic and lepto-hadronic scenarios \cite{Cerruti:2012}. Then, a more comprehensive campaign was conducted from June to October 2013, involving NuSTAR, H.E.S.S., the Swift Observatory and Fermi-LAT \cite{HESS:2020}. For the first time, contemporaneous observations across a broad energy spectrum, from ultraviolet to TeV \(\gamma\)-rays, provided improved X-ray and \(\gamma\)-ray coverage compared to previous campaigns.
Multi-wavelength data from the 2013 campaign have been used to study source variability and X-ray spectra, exploring lepto-hadronic modeling scenarios for the neutrino emission and investigating the low-state period of the source. 

\subsection{Model optimization}
The best-fit LeHa-Paris parameters reproducing the observations were determined through a \(\chi^2\) optimization starting from values outlined in \cite{HESS:2020}, exploring the parameter space of the electron normalization \(K_e\), the proton-to-electron ratio at Lorentz factor \(\gamma=1\) (\(\eta\)), the proton spectral index \(\alpha_p\), the proton normalization at \(\gamma_p^* = 10^4\) (\(K_p^*\)), the emission region size \(R\) and magnetic field \(B\), the electron energy break \(\gamma_{\text{break}}\), and maximum proton energy \(\gamma_p^{\max}\). A Doppler factor \(\delta=33\)~\cite{HESS:2020} and line-of-sight angle \(0.1^\circ\) \cite{Kovalev:2025kxf} are assumed and imply a bulk Lorentz factor \(\Gamma \sim 16\) for the emitting region, while the slopes of the primary electrons energy distribution and their minimum and maximum Lorentz factors are fixed to the values reported in \cite{HESS:2020}: $\alpha_{e,1} = 2.5, \alpha_{e,2}=4.6, \gamma_{min}=2 \cdot 10^3,\gamma_{max}=10^7$. Optimizing over \(K_p^*\) rather than the proton normalization at \(\gamma=1\) enables effective exploration of different \(\alpha_p\), since low-energy protons minimally affect Bethe-Heitler or photo-meson processes. \(K_p^*\) at \(\gamma_p^*=10^4\) better constrains the high-energy proton distribution, with the relation:
\[
K_p(\gamma^*) = \eta(\gamma=1) K_e(\gamma=1) (\gamma^*)^{-\alpha_p},
\]
allowing retrieval of \(\eta(\gamma=1)\) from \(K_p^*\).
The normalization \( K_p^* \) increases for lower values of \( \gamma_p^{\text{max}} \), as a less extended proton energy distribution requires a higher normalization to satisfy observational constraints.
To avoid exploring vast parameter spaces repeatedly, specific ranges for $K_p^*$ have been defined depending on the $\gamma_p^{\text{max}}$ under study, as detailed in Table \ref{pippo}.
\begin{table}[!ht]
\centering
\begin{tabular}{@{}lc@{}}
\toprule
\(\log_{10} \gamma_p^{\max}\) & Range of \(K_p^{\star}\) \\
\midrule
5.0 & \([K_p^{\star} \times 10^{3.2},\; K_p^{\star} \times 10^{3.6}]\) \\
5.6 & \([K_p^{\star} \times 10^{1.8},\; K_p^{\star} \times 10^{2.2}]\) \\
6.2 & \([K_p^{\star} \times 10^{1.0},\; K_p^{\star} \times 10^{1.4}]\) \\
6.8 & \([K_p^{\star} \times 10^{0.2},\; K_p^{\star} \times 10^{0.6}]\) \\
7.4 & \([K_p^{\star} \times 10^{-0.2},\; K_p^{\star} \times 10^{0.2}]\) \\
8.0 & \([K_p^{\star} \times 10^{-0.6},\; K_p^{\star} \times 10^{-0.2}]\) \\
\bottomrule
\end{tabular}
\caption{Ranges of the proton normalization \(K_p^{\star}\) for each value of \(\log_{10} \gamma_p^{\max}\).}
\label{pippo}
\end{table}

\(\alpha_p\) values have been explored discretely as \(\{1.8, 2.0, 2.2\}\), while the other parameters are sampled on a grid of 6 equally spaced points within the following ranges: \(\log_{10} \gamma_p^{\max} \in [5, 8]\), \(K_e \in [10^{3.7}, 10^{4.7}]\) cm\(^{-3}\), \(B \in [9 \times 10^{-3}, 5 \times 10^{-2}]\) G, \(R \in [10^{16.5}, 10^{17}]\) cm, and \(\gamma_{\mathrm{break}} \in [10^{4.6}, 10^{5.1}]\). All the 139968 scenarios have been explored and their models have been compared to the observed electromagnetic SED through a \(\chi^2\) statistic. The reduced \(\chi^2\) has been computed as \(\chi^2/(N-1)\), where $N$ is the number of data points.\\
The best-fit solutions minimize the reduced \(\chi^2\) while also satisfying additional physical constraints. In particular, the radio data serve as a strong upper limit, and any models that exceed this constraint are discarded. Furthermore, among the acceptable solutions, preference is given to those that minimize the total jet luminosity. After computing the \(\chi^2\) grid for all parameters combinations, for each fixed \(\gamma_p^{\max}\), the solution with the lowest reduced \(\chi^2\) was selected. The solutions corresponding to the two highest values of \(\gamma_p^{\max}\) have similar energy budgets ($1.9 \ 10^{48}$ erg/s and $1.3 \ 10^{48}$ erg/s respectively), and the lowest jet power in the sample. As is well known in blazar hadronic modeling, the total power required to produce the observed emission is a critical parameter, as it can easily exceed the Eddington luminosity of the black hole by orders of magnitude. In this case, the solutions with the lowest power are still super-Eddington, but remain within a factor of around 10. For reference, the Eddington luminosity is \(L_{\text{Edd}} \simeq 10^{47}~\text{erg s}^{-1}\) for a supermassive black hole of mass \(M_{\text{BH}} = 10^9~M_\odot\). Among these two solutions, the one at the lowest reduced $\chi^2$ has been selected as optimal solution. Best-fit parameters are reported in Table \ref{tab:paropt}. 
\begin{table}[h!]
\centering
\begin{tabular}{@{}llllllllll@{}}
\toprule
\textbf{$K_e$} [cm$^{-3}$]& \textbf{log$_{10}(\gamma_p^{\text{max}}$}) & \textbf{$\eta$} & \textbf{$\alpha_p$} & \textbf{$K_p^*$} [cm$^{-3}$]& \textbf{B} [G] & \textbf{R} [cm] & \textbf{$\gamma_{\text{break}}$} & \textbf{Reduced $\chi^2$} \\ \midrule
1.3 $10^4$ & 7.4 & 0.003 & 1.8 & 2.3 $10^{-6}$ & 0.036 & 7.9 $10^{16}$ & 6.3 $10^4$ & 2.08\\ \bottomrule
\end{tabular}
\caption{Parameters of the best-fit model.}
\label{tab:paropt}
\end{table}

\begin{figure}[h!]
    \centering
    \includegraphics[width=0.68\linewidth]{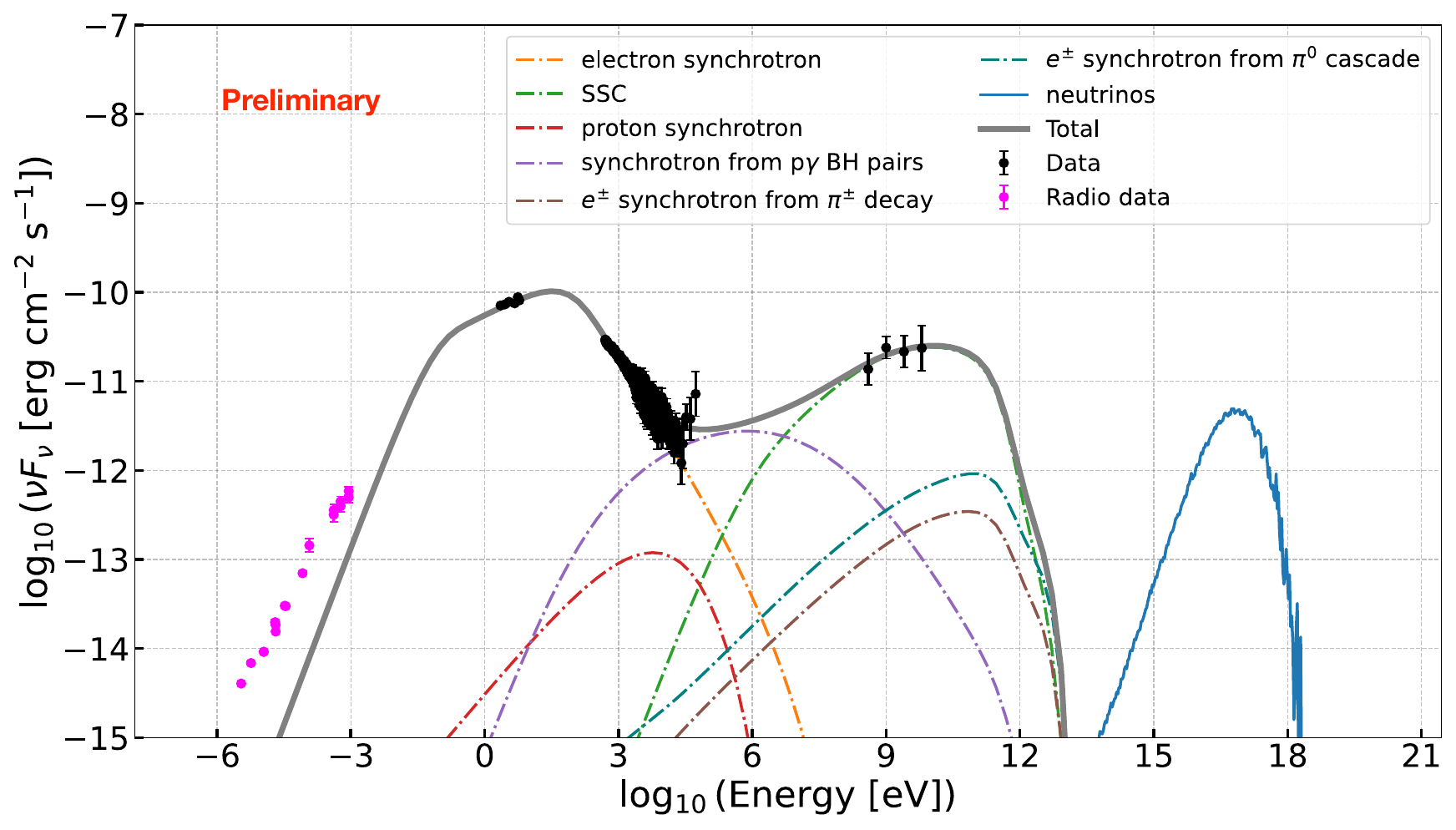}
    \caption{SED of PKS 2155-304 for the best-fit model, compared to data from \cite{Madejski:2016evb} and \cite{HESS:2020}. Associated parameters are reported in Table \ref{tab:paropt}. The all-flavor neutrino SED is reported as a blue solid line. Radiation components not shown in the plot are subdominant.}
    \label{fig:SEDbest}
\end{figure}

Figure \ref{fig:SEDbest} reports the SED modeled with the best-fit parameters: the optimal solution requires an acceleration mechanism for the protons with an $E^{-1.8}$ spectrum, while the resulting size is larger with respect to what was claimed in \cite{HESS:2020} with a less intense magnetic field at the emitting region. The $\gamma_{\text{break}}$ value reflects the position of the first bump of the SED. The all-flavor neutrino SED exhibits a peak at the level of $100$ PeV.

\section{Extension to astrophysical catalogues}
As mentioned earlier, the aim of this work is to extend the neutrino flux optimization done on PKS 2155-304 to an astrophysical catalogue of blazars. In particular, this contribution focuses on the 3HSP catalogue \cite{Chang:2019vfd}, which provides the most extensive list of High Synchrotron Peaked and Extreme HBLs to date. Instead of re-optimizing the model for each source, their theoretical expected neutrino flux can be obtained rescaling the one of PKS 2155-304 according to the methodology described below. \\
For each source, the required measured properties, synchrotron peak flux and redshift, are extracted from the catalogue itself. To rescale the neutrino flux from a reference blazar at redshift \( z \) and synchrotron peak flux \( F_{\gamma, \text{peak}} \) to a target source with redshift \( z' \) and flux \( F_{\gamma, \text{peak}}' \), the following transformations are considered:

\begin{itemize}
    \item \textbf{Energy scaling:} due to the cosmological expansion of the Universe, the observed frequency ($ \nu_{\text{obs}}$) transforms as:
    \[
    \nu'_{\text{obs}} = \nu_{\text{obs}} \cdot \frac{1+z}{1+z'},
    \]
    which is subsequently moved to an energy via the Planck constant.

    \item \textbf{Flux scaling:} working under the assumption that all sources are intrinsically similar in terms of their physical conditions, namely that particle acceleration and radiation mechanisms are comparable across the population, and that external photon fields do not contribute to neutrino production, it is reasonable to adopt the synchrotron peak flux \( F_{\gamma,\text{peak}} \) as a proxy for the neutrino flux ($F_{\nu}$), which transforms to $F_{\nu}'$ according to:
    \[
     F_{\nu}^{'} = F_{\nu} \cdot \frac{F_{\gamma,peak}'}{F_{\gamma,peak}}
    \]

\end{itemize}

This method preserves the spectral shape of neutrino templates across the  while adjusting their normalization based on synchrotron flux. A relevant class of sources for applying and validating this procedure is a subset of HBLs from the 3HSP catalogue. 

\subsection{Model validation}
\begin{figure}[h!]
    \centering
    \includegraphics[width=0.68\linewidth]{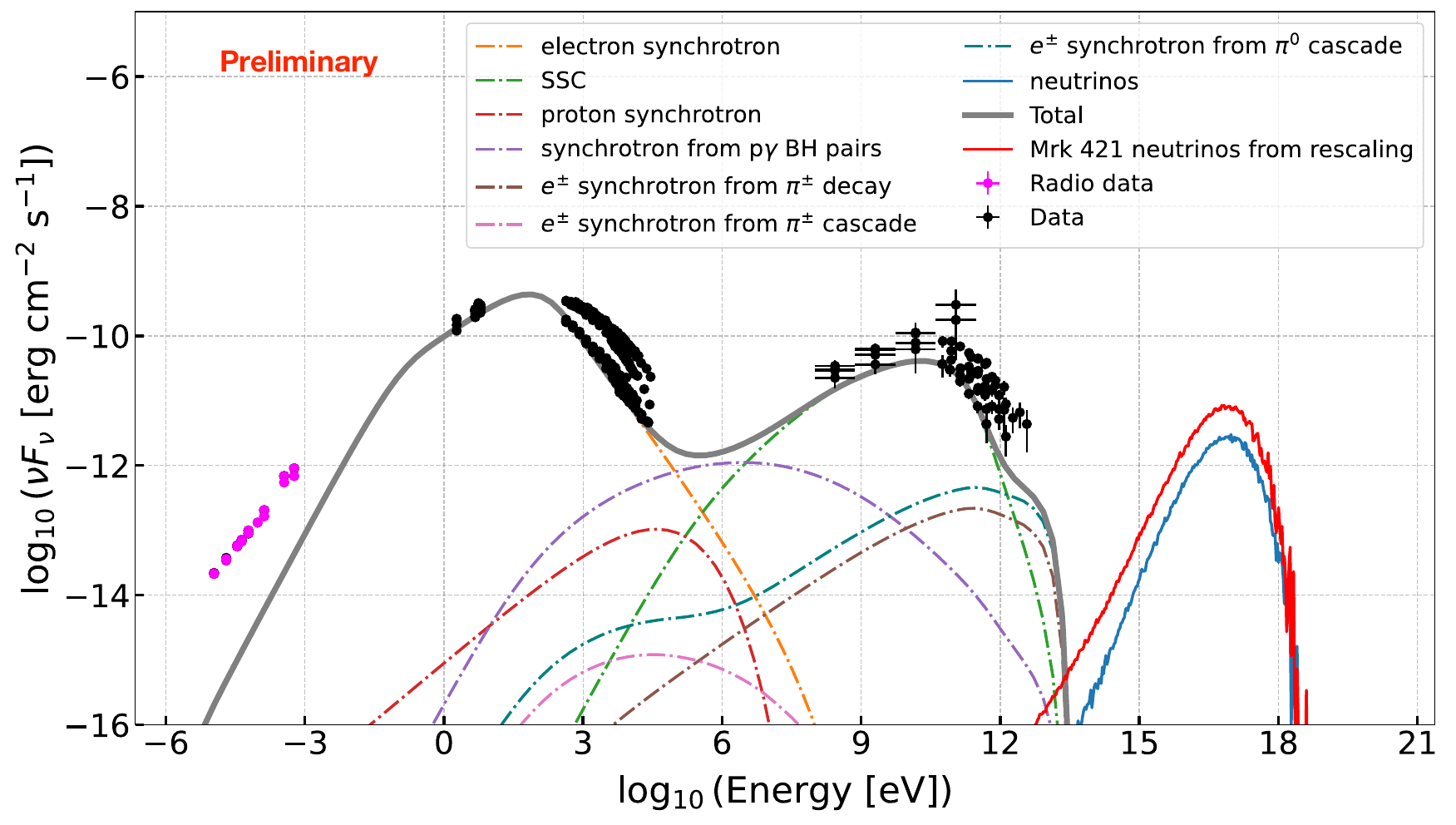}
    \caption{Mrk 421 SED based on parameters reported in Table~\ref{tab:parmrk}. The light-blue line shows the neutrino emission predicted by the model; the red line shows the PKS 2155$-$304 template rescaled to Mrk 421. Radiation components not shown in the plot are subdominant. Data are taken from \cite{Balokovic:2016}.}
    \label{fig:mrksed}
\end{figure}

\begin{figure}[h!]
    \centering
    \includegraphics[width=0.68\linewidth]{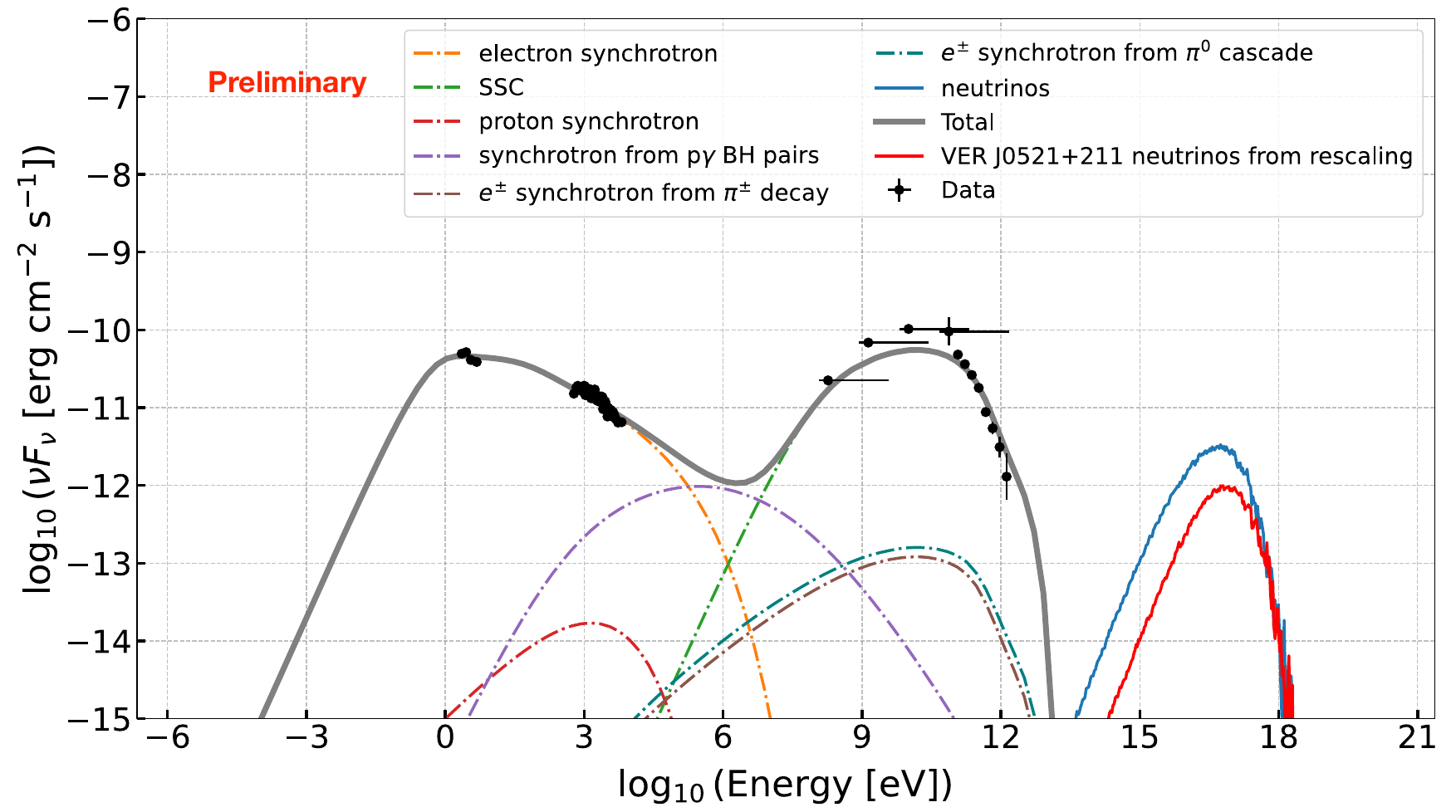}
    \caption{SED of VER J0521+211 based on parameterers reported in Table~\ref{tab:parver}. The blue solid line is the neutrino spectrum of the model, while the solid red corresponds to PKS 2155$-$304 template rescaled to this source. Radiation components not shown in the plot are subdominant. Data are taken from \cite{VERITAS_MAGIC:2022}.}
    \label{fig:versed}
\end{figure}

The phenomenological approach has been validated considering as target sources the HBLs Mrk 421 \cite{Balokovic:2016} and VER J0521+211 \cite{VERITAS_MAGIC:2022}, well-studied in multi-wavelength observational campaigns and exhibiting high synchrotron peak fluxes. The same proton content as PKS 2155$-$304 was assumed for both sources, while allowing the parameters of the electron population and source properties (magnetic field $B$ and emission region size $R$) to vary.\\
Mrk 421 is a nearby HBL located at $z = 0.03$, known for strong variability across the electromagnetic spectrum. Its SED typically shows a synchrotron peak in the X-ray band, shifting toward higher energies during flaring states. The source is often used as a benchmark for emission model testing due to its brightness (log$_{10}(\nu F_{\gamma,\text{peak}}~[\text{erg}\,\text{cm}^{-2}\,\text{s}^{-1}]) = -9.7$). Broadband SED data were taken from simultaneous observations reported in \cite{Balokovic:2016}. These were used to construct a lepto-hadronic model with LeHa-Paris, focusing on the low-activity state to ensure a conservative match. The resulting parameters are reported in Table~\ref{tab:parmrk}, and the corresponding SED is shown in Figure~\ref{fig:mrksed}, where the neutrino emission derived from the model is presented along with the PKS 2155$-$304 neutrino template rescaled to Mrk 421.\\
VER J0521+211 is a TeV HBL located at redshift $z = 0.108$ (as reported in the 3HSP catalogue). The source is among the brightest in the sample in terms of synchrotron peak flux
\begin{table}[h!]
\centering
\begin{tabular}{@{}llllllll@{}}
\toprule
\textbf{B} [G]& \textbf{R [cm]} & \textbf{$\eta$} & \textbf{$K_e$ [cm$^{-3}$]} & \textbf{$\alpha_{e,1}$} & \textbf{$\alpha_{e,2}$} & \textbf{$\gamma_{\text{break}}$} \\ \midrule
0.2 & 0.9 $10^{16}$& 0.06 & 600 & 2.0 & 5.0 & 3 $10^4$\\ \bottomrule
\end{tabular}
\caption{Parameters of the Mrk 421 model. Parameters not specified in this table are the same of PKS~2155-304 best-fit model. Resulting jet luminosity is $3 \cdot 10^{46}$ erg/s.}
\label{tab:parmrk}
\end{table}
(log$_{10}(\nu F_{\gamma,\text{peak}} ~[\text{erg}\,\text{cm}^{-2}\,\text{s}^{-1}]) = -10.6$). It has been extensively monitored in observational campaigns, notably during an elevated TeV state between October and December 2013 \cite{VERITAS_MAGIC:2022}. A lepto-hadronic model was developed for this source using LeHa-Paris, under the same assumptions as before. The best-fit parameters are summarized in Table~\ref{tab:parver}. The resulting SED is shown in Figure~\ref{fig:versed}, where the modeled neutrino flux is presented along with the PKS 2155$-$304 neutrino template rescaled to match VER J0521+211. 
 
\newpage
\begin{table}[h!]
\centering
\begin{tabular}{@{}llllllll@{}}
\toprule
\textbf{$B$ [G]} & \textbf{$R$ [cm]} & \textbf{$\eta$} & \textbf{$K_e$ [cm$^{-3}$]} & \textbf{$\alpha_{e,1}$} & \textbf{$\alpha_{e,2}$} & \textbf{$\gamma_{\text{break}}$} \\ \midrule
0.01 & $1.3 \ 10^{17}$ & $3 \ 10^{-6}$ & $10^7$ & 3.0 & 3.6 & $10^5$ \\ \bottomrule
\end{tabular}
\caption{Parameters of the VER~J0521+211 model. Parameters not listed are identical to those of the PKS~2155$-$304 best-fit model. Resulting jet luminosity is $3 \cdot 10^{48}$~erg/s.}
\label{tab:parver}
\end{table}
The ratio between the modeled neutrino fluxes and the rescaled PKS 2155-304 flux shows that, in the neutrino energy range \(1~\mathrm{PeV} \leq E_\nu \leq 100~\mathrm{PeV}\), the Mrk 421 model agrees with the rescaling within a factor of 5, while the VER J0521+211 model agrees within a factor of 10.
\section{Conclusions and perspectives}
Starting from a re-optimization of the spectral models of Mrk 421 and VER J0521+211, the obtained results are consistent with those derived from rescaling the PKS 2155-304 template, confirming the robustness of the adopted approach. By employing the LeHa-Paris code, it is possible to generate expected neutrino spectra for a specific subclass of blazars. The good agreement between the modeled fluxes and the rescaled template, especially in the PeV energy range and beyond, makes these studies particularly relevant for neutrino telescopes. Looking ahead, integrating such theoretical models into stacking analyses using data from next-generation deep-sea Cherenkov neutrino telescopes, such as KM3NeT/ARCA, offers a promising path to enhance our understanding of high-energy neutrino sources, as highlighted in~\cite{careniniproc}.

\begingroup
\fontsize{11}{11}\selectfont

\endgroup

\begin{thebibliography}{99}

\bibitem{IceCube:2018dnn}
M.~G. Aartsen \textit{et al.}, Science \textbf{361} (2018) no.6398, eaat1378

\bibitem{Cerruti:2014iwa}
M. Cerruti \textit{et al.}, Mon. Not. Roy. Astron. Soc. \textbf{448}, no.1, 910-927 (2015)

\bibitem{Cerruti:2019mnras}
M. Cerruti \textit{et al.}, MNRAS Lett. \textbf{483}, no.1, L12–L16 (2019)

\bibitem{KM3Net:2016zxf}
S.~Adrián-Martinez \textit{et al.}, J. Phys. G \textbf{43}, no.8, 084001 (2016)

\bibitem{Mucke2000}
A. M\"ucke \textit{et al.}, 2000, Computer Physics Communications, \textbf{124}, 290

\bibitem{Petropoulou2015}
M. Petropoulou \textit{et al.}, 2015, Mon. Not. Roy. Astron. Soc., \textbf{447}(1), 36--48

\bibitem{2024arXiv241114218C}
M. Cerruti \textit{et al.}, arXiv:2411.14218 (2024)

\bibitem{Madejski:2016evb}
G.~M. Madejski \textit{et al.}, Astrophys.\ J.\ \textbf{831} (2016) no.2, 142

\bibitem{Cerruti:2012}
M. Cerruti \textit{et al.},  
PoS \textbf{Gamma2012} (2013) 040

\bibitem{HESS:2020}
H. Abdalla \textit{et al.}, Astron. Astrophys. \textbf{639} (2020) A42

\bibitem{Kovalev:2025kxf}
Y.~Y. Kovalev \textit{et al.},
arXiv:2504.09287 (2025)

\bibitem{Chang:2019vfd}
Y.~L. Chang \textit{et al.}, Astron. Astrophys. \textbf{632}, A77 (2019)

\bibitem{Balokovic:2016}
M. Baloković \textit{et al.}, Astrophys. J. \textbf{819} (2016) no.2, 156

\bibitem{VERITAS_MAGIC:2022}
C.~B. Adams \textit{et al.}, Astrophys. J. \textbf{932} (2022) no.2, 129

\bibitem{careniniproc}
F. Carenini \textit{et al.}, PoS \textbf{ICRC2025} (2025) 1005


\end{thebibliography}
\end{document}